\documentclass[10pt,letterpaper,twoside,fleqn]{article}
\usepackage{amsmath}
\usepackage{amssymb}
\usepackage{amstext}
\usepackage{theorem}
\usepackage{mathrsfs}
\usepackage{graphics}
\usepackage{ulem}
\usepackage{moredefs}
\usepackage[mla]{lips}
\usepackage{csquotes}
\usepackage[pdftex]{graphicx}
\usepackage{caption}

\numberwithin{equation}{section}
\numberwithin{figure}{section}

\newtheorem{theorem}{Theorem}[section]

\newtheorem{corollary}[theorem]{Corollary}

\newtheorem{proposition}[theorem]{Proposition}

\newtheorem{defi}[theorem]{Definition}

\newcommand{\nn}{\nonumber}

\newcommand{\unity}{{\setlength{\unitlength}{1em}
\begin{picture}(0.75,0.75)
\put(0,0){$1$} \put(.38,0){\line(0,1){0.65}}
\end{picture}}}

\begin{document}

\title{\bf Instability of pre-existing resonances in the DC Stark effect vs. stability in the AC Stark effect}
\author{I. Herbst\footnote{University of Virginia, Department of Mathematics, 141 Cabell Drive, Charlottesville, VA 22903, USA} \and
J. Rama\footnote{Universit\"at Potsdam, Institut f\"ur Mathematik, Am Neuen Palais 10, 14469 Potsdam, Germany}}
\date{\small March 29, 2014}
\maketitle

\begin{abstract}
\noindent For atoms described by a 1-dimensional Friedrichs model we present the following result: pre-existing resonances, when subjected to a small constant electric field (DC Stark effect), are unstable in the weak field limit. In contrast, pre-existing resonances under the influence of a small time-periodic electric field (AC Stark effect) are stable in this limit. This article is a review of our results in \cite{HeRa}.
\end{abstract}

\section{Introduction}

It is well known that atomic bound states below the continuum turn into resonances under the influence of a weak constant electric field (DC Stark effect). These resonances are well defined in a translation or dilation analytic framework and move continuously as a function of the field strength, $f$, for small $|f|$ and converge to the original bound state in the limit $f\to 0$; see, e.g., \cite{CRA}, \cite{GG1}, \cite{GG2}, \cite{HaSi}, \cite{He}, \cite{HeSi1}, \cite{HeSi2},  \cite{R}, \cite{YTS} and references given there. (In this sense eigenvalues below the continuous spectrum are stable under the electric DC field.) It is natural to ask: Is the same true for (nonreal) pre-existing resonances $r_0$ in atoms (which might arise as shape resonances or due to a broken symmetry)? In this paper we discuss and answer this question in a very simple model, namely the Friedrichs model.  Complete proofs of our results can be found in \cite{HeRa}.

At first glance one might expect that essentially the same methods as used in the case of bound states turning into resonances should be sufficient to treat this problem. However, our somewhat surprising result is that this is not true: Not only do these methods fail, but the expected stability result itself does not hold, at least in the physically most relevant case of pre-existing resonances near the real axis.
This is due to the blow-up of relevant resolvents as $f\to0$ for values of the spectral parameter near the pre-existing original resonance; see \cite{He}.

We emphasize that this instability of resonances (if they are sufficiently isolated and near the real axis) should be an observable effect in the laboratory.

This instability result in the DC Stark effect (with electric field $fx$) is in marked contrast to the AC Stark effect (with time dependent electric field $fx\sin(\omega t)$). Here, in the dilation analytic framework, resonances for the time-dependent family of Hamiltonians $H(t,f)$ (as $t$ varies over a period $[0,2\pi/\omega]$) are well defined in an adaptation of the Howland-Yajima formalism (see \cite{CyFKS}, \cite{How}, \cite{Y2}, \cite{Y}) to our Friedrichs model, as nonreal eigenvalues of the dilated Floquet Hamiltonian $K(\theta,f)$ for nonreal $\theta$ (see \cite{HeRa}).

This dichotomy of instability versus stability is reminiscent of the following situation in classical mechanics: Introducing a periodic change in parameters may turn an unstable equilibrium into a stable one. (See for example the unstable equilibrium for a pendulum; \cite{Arnold}.)

Our results need dilation analyticity or analyticity in momentum space in the following sense: Let
$S_{\theta_0}:=\{z\in\mathbb{C}\,|\,|\textrm{Im\,}z|<\theta_0\}$ $(\theta_0>0)$.
\begin{defi}\label{defi:dilation_analytic_vector}{\rm(cf. \cite[Definition 1.1]{HeRa})}
\begin{align}\hspace{-1cm}\textrm{{\rm(i)}}\hspace{2cm} (U(\theta)\psi)(x):= e^{\theta/2}\psi(e^{\theta}x)\hspace{0.5cm}(\theta\in\mathbb{R},\ \psi\in L^2(\mathbb{R}))\label{unitary_group}\end{align} defines the unitary group of dilations on $L^2(\mathbb{R})$. A function $\psi$ is called  dilation analytic (in angle $\theta_0$), if $U(\theta)\psi$ has an $L^2$-valued analytic extension (in the variable $\theta$) to a strip $S_{\theta_0}$ for some $\theta_0>0$.   $\mathcal{D}_{\theta_0}$ denotes the linear space of $L^2$-functions which are dilation analytic in angle $\theta_0$.\smallskip\\
\begin{align}\hspace{-1cm}\textrm{{\rm(ii)}}\hspace{2cm}
\mathcal{T}_{k_0}:=\big\{\psi\in L^2(\mathbb{R})\,\big|&\,\widehat{\psi} \textrm{ (the Fourier transform of $\psi$) has an analytic}\nn\\
&\ \textrm{extension to the strip }S_{k_0}
\textrm{ and }\sup\limits_{k\in S_{k_0}}|\widehat{\psi}(k)|<\infty\big\},\nn\end{align} for all $k_0>0$.
\end{defi}

In the following we write $(\,\cdot\,,\,\cdot\,)_X$ for an inner product in a linear space $X$ and use the notation $\mathbb{C}_\pm:=\{z\in\mathbb{C}\,|\,\textrm{Im\,}z\gtrless0\}$. $\overline{\mathbb{C}_\pm}$ denotes the closure of $\mathbb{C}_\pm$. By $\sqrt{\cdot}$ we mean the principal branch of the square root with branch cut $(-\infty,0)$.

The outline of this article is as follows: In Section \ref{section:2} we describe our central result on instability of pre-existing resonances in the DC Stark effect. In Section \ref{section:4} we report on numerical calculations
verifying this analytic result on instability. In particular, there seems to be a cloud of resonances around the pre-existing resonance, $r_0$, converging to the real axis as $f\downarrow0$. In Section \ref{section:3} we describe stability of pre-existing resonances in the AC Stark effect.

\section{Friedrichs model with DC Stark effect}\label{section:2}

Let $p$ denote the self-adjoint realization of $-i d/dx$ in $L^2(\mathbb{R})$. Given the Hilbert space $\mathscr{H}:=L^2(\mathbb{R})\oplus\mathbb{C}$ with inner product
\begin{align} (u,u)_{\mathscr{H}}:=(u_1,u_1)_{L^2}+(u_2,u_2)_{\mathbb{C}}\hspace{0.5cm}(u:=\begin{pmatrix} u_1\cr u_2\end{pmatrix}\in L^2(\mathbb{R})\oplus\mathbb{C}),\label{inner_product}\end{align}
the self-adjoint realization in $\mathscr{H}$ of
\begin{align} H_0:=\begin{pmatrix}p^2 & 0 \cr 0 & 1\cr \end{pmatrix}\,,
\label{H_0}\end{align} has a simple eigenvalue, 1, embedded in its essential spectrum $\sigma_{ess}(H_0)=[0,\infty)$. Then adding a small rank-2 perturbation gives the Hamiltonian
\begin{align} H_\varphi:=\begin{pmatrix}p^2 & \varphi \cr (\varphi\,,\,\cdot\,)_{L^2} & 1 \cr\end{pmatrix}\hspace{0.5cm}(\varphi\in\mathcal{D}_{\theta_0}\cup\mathcal{T}_{k_0}\textrm{ for some $k_0,\theta_0>0$})\,,\label{H}\end{align} where $\theta_0$ and $k_0$ depend (in a certain sense specified below) on the size of $\varphi$, and  $\varphi$ in the right upper corner of $H_\varphi$ is identified with the multiplication operator generated by the function $\varphi$. This perturbation removes the embedded eigenvalue of $H_0$ and turns it into a nearby resonance of $H_\varphi$ (in the sense of Definition \ref{defi:resonance} below). Adding an external electric field the Hamiltonian is modeled by \begin{align} H_\varphi(f):=\begin{pmatrix} p^2+f x & \varphi \cr (\varphi\,,\,\cdot\,)_{L^2} & 1 \cr\end{pmatrix}\hspace{0.5cm}(x\in\mathbb{R},\ f\geq0,\ \varphi\in\mathcal{D}_{\theta_0}\cup\mathcal{T}_{k_0})\,.\label{H(f)}\end{align}
\noindent By standard arguments, the operators $H_0$ and $H_\varphi(f)$ $(f\geq0)$, $H_\varphi(0)=H_\varphi$ are essentially self adjoint on $C_0^{\infty}(\mathbb{R})\oplus\mathbb{C}$. We also use a standard definition of resonances as poles of the meromorphic extension of certain matrix elements of the resolvent of $H_0$ and $H_\varphi(f)$. More precisely:
\begin{defi}\label{defi:resonance}{\rm(\cite[Definition 1.2]{HeRa})}\\
Let $\kappa_0>0$ and $0<\vartheta_0<\pi/3$.  Assume {\rm\eqref{inner_product}} -- {\rm\eqref{H(f)}}.
Fix $\varphi\in\mathcal{D}_{\theta_0}\cup\mathcal{T}_{k_0}$, where $k_0\geq\kappa_0$ and $\theta_0\geq\vartheta_0$. Let \begin{align} \Omega_f:=\left\{\begin{array}{l@{\,,\quad }l}\{z\in\mathbb{C}\,|\,z=|z|e^{i\theta},\ 0\leq|\theta|<2\vartheta_0\}\cap\mathbb{C}_-&\varphi\in\mathcal{D}_{\theta_0}\textrm{ and }f=0\\ S_{\kappa_0}\cap\mathbb{C}_-&\varphi\in\mathcal{T}_{k_0}\textrm{ and }f=0\\
\mathbb{C}_- & \varphi\in\mathcal{D}_{\theta_0}\cup\mathcal{T}_{k_0}\textrm{ and }f>0\end{array}\right.\,.\nn\end{align}
\begin{itemize}
\item[{\rm (i)}] Let $f\geq0$. A number $z_*$ in $\mathbb{C}_-$ is defined to be a resonance of $H_\varphi(f)$, if for some $\Psi:=\begin{pmatrix}\psi \cr c \end{pmatrix}\in L^2(\mathbb{R})\oplus\mathbb{C}$, with $\psi\in\mathcal{D}_{\vartheta_0}$ if $\varphi\in\mathcal{D}_{\theta_0}$ and $\psi\in\mathcal{T}_{\kappa_0}$ if $\varphi\in\mathcal{T}_{k_0}$, the meromorphic extension of the resolvent matrix element $(\Psi\,,\,(H_\varphi(f)-z)^{-1}\Psi)_{\mathscr{H}}$ $(z\in\mathbb{C}_+)$ to the region $\Omega_f\cup\overline{\mathbb{C}_+}\backslash(-\infty,0]$ has a pole at $z_*$.
\item[{\rm (ii)}] For any function $g_{f,\varphi}(\cdot)$ (depending on given $\varphi$ and $f$) which is analytic in $\mathbb{C}_+$ the symbol $g_{f,\varphi}^{\mathbf{c},\Omega_f}(\cdot)$ denotes the analytically (or meromorphically) continued function to the region $\Omega_f\cup\overline{\mathbb{C}_+}\backslash(-\infty,0]$.
\end{itemize}
\end{defi}
The region $\Omega_f$ $(f\geq0)$ is a region of analyticity of $(u\,,\,(p^2+fx-\,\cdot\,)^{-1}v)_{L^2}^{\mathbf{c},\Omega_f}$ for all $u$ and $v$ in $\mathcal{D}_{\theta_0}$ or in $\mathcal{T}_{k_0}$. Setting, for $z\in\mathbb{C}_+$, \begin{align} R_f(z):=(p^2+fx-z)^{-1}\,,\quad F_{f,\varphi}(z):=1-z-(\varphi\,,\,R_f(z)\varphi)_{L^2}\label{F_f}\end{align}
the resolvent of the matrix operator $H_\varphi(f)$ is given by
\begin{align} (H_\varphi(f)-z)^{-1}=\begin{pmatrix}R_f(z)\big(1 +F_{f,\varphi}(z)^{-1}(\varphi\,,\,R_f(z)\,\cdot\,)_{L^2}\varphi\big) & -F_{f,\varphi}(z)^{-1}M_{R_f(z)\varphi}\cr -F_{f,\varphi}(z)^{-1}(\varphi\,,\,R_f(z)\,\cdot\,)_{L^2} & F_{f,\varphi}(z)^{-1}\cr\end{pmatrix},\nn
\end{align} where $z\in\mathbb{C}_+$, $\varphi\in\mathcal{D}_{\theta_0}\cup\mathcal{T}_{k_0}$ and $f\geq0$. Then for any $\psi\in\mathcal{D}_{\theta_0}\cup\mathcal{T}_{k_0}$ and $c\in\mathbb{C}$ one finds, for all $z\in\mathbb{C}_+$,
\begin{align} (\Psi\,,\,(H_\varphi(f)-z)^{-1}\Psi)_{\mathscr{H}} = & \ (\psi,R_f(z)\psi)_{L^2}+F_{f,\varphi}(z)^{-1}(\varphi,R_f(z)\psi)_{L^2}(\psi,\varphi)_{L^2}\nn\\
&\ -c F_{f,\varphi}(z)^{-1}(\psi,R_f(z)\varphi)_{L^2} -\overline{c}F_{f,\varphi}(z)^{-1}(\varphi,R_f(z)\psi)_{L^2}\nn\\
&\ +|c|^2 F_{f,\varphi}(z)^{-1}. \label{2.6}\end{align}
Then continuing from $\mathbb{C}_+$ to $\Omega_f\cup\overline{\mathbb{C}_+}\backslash(-\infty,0]$ shows that the only way poles in $\Omega_f$ of $(\Psi\,,\,(H_\varphi(f)-\,\cdot\,)^{-1}\Psi)_{\mathscr{H}}^{\mathbf{c},\Omega_f}$ can arise is as zeros in $\Omega_f$ of the function $F_{f,\varphi}^{\mathbf{c},\Omega_f}$. \eqref{2.6} also shows that our Definition \ref{defi:resonance} of a resonance is in fact independent of the choice of the test function $\Psi$. This proves
\begin{proposition}\label{proposition:resonance=solution}{\rm(\cite[Proposition 1.4]{HeRa})}\\
Let $\theta_0$ and $k_0$ be as in Definition \ref{defi:resonance}. Let $\varphi\in\mathcal{D}_{\theta_0}\cup\mathcal{T}_{k_0}$, $f\geq0$. Let $H_\varphi(f)$ and $F_{f,\varphi}$ be given by {\rm\eqref{H(f)}} and {\rm\eqref{F_f}}. The resonances of $H_\varphi(f)$ are precisely the solutions of
$F_{f,\varphi}^{\mathbf{c},\Omega_f}(z)=0$ in $\Omega_f$ (using the notation of Definition \ref{defi:resonance}). The real zeros of $F_{f,\varphi}^{\mathbf{c},\Omega_f}$ are eigenvalues of the self-adjoint operator $H_\varphi(f)$.
\end{proposition}
Since, for $f>0$ and $u,v\in\mathcal{D}_{\theta_0}$ or $u,v\in\mathcal{T}_{k_0}$, the continued resolvent matrix elements of the 1-dimensional Stark operator, $(u\,,\,(p^2+fx-\,\cdot\,)^{-1}v)_{L^2}^{\mathbf{c},\Omega_f}$, are entire (see \cite{He} and \cite{HeRa}), the function $F_{f,\varphi}(\,\cdot\,)$ has an entire extension, more precisely:

\begin{proposition}\label{proposition:infinitely_many_zeros}{\rm(\cite[Proposition 1.5]{HeRa})}\\
 Let $\theta_0$ and $k_0$ be as in Definition \ref{defi:resonance}. Let $\varphi\in\mathcal{T}_{k_0}\cup\mathcal{D}_{\theta_0}$.
Let $H_\varphi(f)$ and $F_{f,\varphi}$ be given by {\rm\eqref{H(f)}} and {\rm\eqref{F_f}}. Then for all $f>0$ the function $F_{f,\varphi}$ has an extension to an entire function of finite order\footnote{An entire function $g$ is defined to be of finite order if there exist $n\in\mathbb{R}$ and $R>0$ such that for all $z\in\mathbb{C}$ with $|z|\geq R$ one has $|g(z)|\leq e^{|z|^n}$; see, e.g., \cite{BN}.}. This entire extension has infinitely many zeros.
\end{proposition}
A consequence of Proposition \ref{proposition:resonance=solution} and  Proposition \ref{proposition:infinitely_many_zeros} is, since $H_\varphi(f)$ has only finitely many real eigenvalues,
\begin{corollary}\label{corollary:infinitely_many_resonances}{\rm(\cite[Corollary 1.6]{HeRa})}\\
 Let $\theta_0$ and $k_0$ be as in Definition \ref{defi:resonance}. Let $\varphi\in\mathcal{T}_{k_0}\cup\mathcal{D}_{\theta_0}$. Suppose in addition that $\varphi$ is in the domain of $\langle p\rangle^{(1+\varepsilon)/2}$ for some $\varepsilon>0$ (as an operator in $L^2(\mathbb{R})$), where $\langle p\rangle:=(1+|p|^2)^{1/2}$.
Then for all $f>0$ the operator $H_\varphi(f)$ has infinitely many resonances.
\end{corollary}

In this model the existence of pre-existing resonances (i.e., resonances of $H_\varphi(0)$) is shown by the following proposition.
\begin{proposition}\label{proposition:existence_resonance}{\rm(\cite[Proposition 1.7, 3.]{HeRa})}\\
 Let $\theta_0$ and $k_0$ be as in Definition \ref{defi:resonance}. Let $\varphi\in\mathcal{T}_{k_0}\cup\mathcal{D}_{\theta_0}$. If $\widehat{\varphi}(\pm\sqrt{\lambda})\neq0$ for $\lambda\in[1-\varepsilon\,,\,1+\varepsilon]$ with some $\varepsilon\in(0,1)$, then for sufficiently small $\varphi$ the operator $H_\varphi(0)$ has exactly one resonance close to 1. The smallness of $\varphi$ is measured by the size of $\varrho^{-1}|(\varphi\,,\,(p^2-z)^{-1}\varphi)^{\mathbf{c},\Omega_0}_{L^2}|$  for $|z-1|=\varrho$, where $\varrho$ is some number in $(0,\varepsilon)$.
\end{proposition}
This proposition follows from Proposition \ref{proposition:resonance=solution} by use of Rouch\a'e's Theorem.

Our central result is that pre-existing resonances of $H_\varphi(0)$ are unstable under the addition of a constant electric field in the following sense:
\begin{theorem}\label{theorem:resonances_instable}{\rm(\cite[Theorem 1.8]{HeRa})}\\
 Let $k_0$ and $\theta_0$ be as in Definition \ref{defi:resonance}.
Let $\varphi\in\mathcal{D}_{\theta_0}\cup\mathcal{T}_{k_0}$. If $\varphi$ is in $\mathcal{D}_{\theta_0}\backslash\mathcal{T}_{k_0}$, suppose in addition that \begin{align} \sup\big\{|\frac{d^n}{dk^n}\,\widehat{\varphi}(e^{i\theta}k)|\,\big|\,k\in\mathbb{R}\backslash\{0\},\ \theta\in\mathbb{R},\ |\theta|<\min\{\theta_0,\frac{\pi}{3}\},\ n\leq2\big\}<\infty\,.\nn\end{align} Let $H_\varphi(f)$, $f>0$, be given by {\rm\eqref{H(f)}}.
If $\varphi\in\mathcal{D}_{\theta_0}\backslash\mathcal{T}_{k_0}$ let
\begin{align} M:=\{z\in\mathbb{C}_-\,|\, 0>{\rm arg\,}\sqrt{z}\geq -\min\{\frac{\pi}{3},\theta_0\}+\varepsilon_1\,,\ {\rm Re\,}\sqrt{z}\in[\varepsilon_2,a]\}\nn\end{align} and if $\varphi\in\mathcal{T}_{k_0}$ let
\begin{align}   M:=\{z\in\mathbb{C}_-\,|\,& 0>{\rm arg\,}\sqrt{z}\geq-\frac{\pi}{3}+\varepsilon_1,\ {\rm Re\,}\sqrt{z}\in[\varepsilon_2,a],\
 0>{\rm Im\,}\sqrt{z}\geq -k_0+\varepsilon_3\}\nn
\end{align}  for any $\varepsilon_1,\varepsilon_2,\varepsilon_3>0$ sufficiently small and any $a>\varepsilon_2$.
Suppose there exists $\delta>0$ such that for all $z\in M$ $|\widehat{\varphi}(\sqrt{z})\overline{\widehat{\varphi}}(-\overline{\sqrt{z}})|\geq\delta$. Then
\begin{align}  \exists\, c_0>0\ \exists\, f_0>0\ &\textrm{sufficiently small}\ \forall\, f\in(0,f_0]:\nn\\
&r \textrm{ is a resonance of }H_\varphi(f)\textrm{ in }M \Rightarrow |{\rm Im\,}r|\leq c_0 f\,.\nn\end{align}
In particular, ${\rm Im\,}r\to0$ $(f\downarrow0)$. Thus $r$ does not converge to any resonance of $H_\varphi(0)$ as $f\downarrow 0$.
\end{theorem}
The proof of Theorem \ref{theorem:resonances_instable} is long and technical. We refer to \cite{HeRa}. Central ideas of the proof are identifying resonances with solutions (in $\Omega_f$) of
\begin{align}
 0=F_{f,\varphi}^{\mathbf{c},\Omega_f}(z)=1-z-r_{f,\varphi}^{\mathbf{c},\Omega_f}(z)\,,\label{eq_zero}
\end{align}
where  $r_{f,\varphi}(z):=(\varphi\,,\,(p^2+fx-z)^{-1}\varphi)_{L^2}$ $(z\in\mathbb{C}_+)$,
and determining the leading order of the asymptotic expansion as $f\downarrow0$ of $r_{f,\varphi}^{\mathbf{c}}(z)$ for $z\in M$ -- given by explicit pole terms -- with uniform bounds on the remainder. This expansion is obtained by using contour deformation and the method of steepest descent.

\section{Friedrichs model with AC Stark effect}\label{section:3}

Our AC Stark Hamiltonian is
\begin{align} H_\varphi(t,f):=\begin{pmatrix} h(t,f)& \varphi \cr (\varphi\,,\,\cdot\,)_{L^2} & 1 \end{pmatrix}\,,\hspace{0.3cm}h(t,f):=p^2+fx\sin(\omega t)\label{H(f)_AC}\end{align} for $f\geq0$ and $x,t\in\mathbb{R}$, where $\omega>0$ denotes the frequency of the alternating electric field and $\varphi$ is in $\mathcal{D}_{\theta_0}$ for some $\theta_0\in(0,\frac{\pi}{2})$.

In this time-periodic setting, resonances are defined as nonreal eigenvalues of the complex dilated Floquet Hamiltonian $K(f,\theta)$ (given in \eqref{K} below) associated with the time-periodic Hamiltonian $H_\varphi(t,f)$. This is basically Yajima's idea, see \cite{Y}  and \cite{Y2}, where he adapts Howland's formalism, \cite{How} (see also \cite{CyFKS}), to periodic problems.  In particular in \cite{Y} Yajima considers the AC Stark problem with a class of analytic potentials.

We adapt the framework of \cite{Y} to our matrix operator \eqref{H(f)_AC}. We define, for $t\in\mathbb{R}$ and $f\geq0$,
\begin{align} &\widetilde{T}(t,f):\ L^2(\mathbb{R})\oplus\mathbb{C}\to L^2(\mathbb{R})\oplus\mathbb{C}\,,\hspace{0.5cm}
\widetilde{T}(t,f):=\begin{pmatrix}T(t,f) & 0 \cr 0 & 1\end{pmatrix}\,,\nn\\
&T(t,f):= e^{i2f\omega^{-2}\sin(\omega t)p}\,e^{-if\omega^{-1}\cos(\omega t)x}\,.\nn\end{align} (Note that $T(t,f)$ is unitary on $L^2(\mathbb{R})$.) If
$i \partial_t\Psi = H_\varphi(t,f)\Psi$, then
\begin{align}
& i\partial_t(\widetilde{T}(t,f)\Psi)=\widetilde{H}_\varphi(t,f)\widetilde{T}(t,f)\Psi\,,\nn\\
& \widetilde{H}_\varphi(t,f):= (i\partial_t\,\widetilde{T}(t,f))\widetilde{T}(t,f)^{-1}+\widetilde{T}(t,f)H_\varphi(t,f)\widetilde{T}(t,f)^{-1}\nn\end{align}
for all $t\in\mathbb{R}$ and all $f\geq0$. The unitary transformation $\widetilde{T}(t,f)$ is the implementation of a gauge transformation via a unitary transformation on $L^2(\mathbb{R})\oplus\mathbb{C}$; cf. Remark 3 (due to Hunziker) in \cite[Chapter 7.3]{CyFKS}. $\widetilde{T}(t,f)$ transforms the unbounded electric potential to zero, modulo a shift which is constant w.r.t. $x$ (see \eqref{htilde} below). One finds, for $f\geq0$ and $t\in\mathbb{R}$,
\begin{align}
&\widetilde{H}_\varphi(t,f)=\begin{pmatrix}\widetilde{h}(t,f) & T(t,f)\varphi \cr (T(t,f)\varphi\,,\,\cdot\,)_{L^2} & 1 \end{pmatrix}\,,\label{HtildeAC}\\
&\widetilde{h}(t,f):= (i\partial_t T(t,f))T(t,f)^{-1}+T(t,f)h(t,f)T(t,f)^{-1}\,,\nn\end{align} where a calculation shows
\begin{align}
\widetilde{h}(t,f)= p^2+\frac{f^2}{2\omega^2}\cos(2\omega t)+\frac{f^2}{2\omega^2}\,.\label{htilde}
\end{align}
Similarly to \cite{Y}, we now consider the so called Floquet Hamiltonian
\begin{align}
K(f):=-i\partial_t + \widetilde{H}_\varphi(t,f)\hspace{0.5cm}(f\geq0\,,\ t\in\mathbb{R})\nn
\end{align} as an operator in
$\mathscr{K}:=(L^2(\mathbb{R})\oplus\mathbb{C})\otimes L^2(\mathbb{T}_\omega)$, where $\mathbb{T}_\omega:=\mathbb{R}/\tau\mathbb{Z}$ with $\tau:=2\pi/\omega$ (the period). We identify $K(f)$ with
\begin{align}
K(f)=\unity_{L^2(\mathbb{R})\oplus\mathbb{C}}\otimes(-i\partial_t)+\widetilde{H}_\varphi(t,f)\otimes\unity_{L^2(\mathbb{T}_\omega)}
\hspace{0.5cm}(f\geq0\,,\ t\in\mathbb{R})\,.\nn
\end{align}  $K(f)$ has a self-adjoint realization in $\mathscr{K}$, which we also denote by $K(f)$. Then $K(f)$ generates a unitary group $\{e^{-isK(f)}\}_{s\in\mathbb{R}}$ in $\mathscr{K}$ which is unitarily equivalent to the propagator
$\widetilde{U}(t+\tau,t;f)\otimes\unity_{L^2(\mathbb{T}_\omega)}$ $(t\in\mathbb{R})$ over the period $\tau$, where
$i\partial_t\widetilde{U}(t,s;f)=\widetilde{H}_\varphi(t,f)\widetilde{U}(t,s;f)$ and $\widetilde{U}(t,t;f)=\unity$ $(s,t\in\mathbb{R})$. (Existence and uniqueness of this propagator follow from results in \cite{How} and \cite{Y}.) We now define resonances by use of dilation analyticity: Let
\begin{align}
V(\theta)^{\pm1}:=\begin{pmatrix}U(\theta)^{\pm1} & 0 \cr 0 & 1 \end{pmatrix}\hspace{0.5cm}(\theta\in S_{\theta_0})\,,\nn
\end{align} where $U(\cdot)$ denotes the group of dilations in $L^2(\mathbb{R})$ (see \eqref{unitary_group}). Set
\begin{align}
K(f,\theta):=-i\partial_t+V(\theta)\widetilde{H}_\varphi(t,f)V(\theta)^{-1}\hspace{0.5cm}(\theta\in S_{\theta_0},\ f\geq0,\ t\in\mathbb{R})\,,\label{K}
\end{align} where
\begin{align} & V(\theta)\widetilde{H}_\varphi(t,f)V(\theta)^{-1}=\begin{pmatrix}U(\theta)\widetilde{h}(t,f)U(\theta)^{-1}
 & U(\theta)T(t,f)\varphi \cr (U(\overline{\theta})T(t,f)\varphi\,,\,\cdot\,)_{L^2} & 1 \end{pmatrix}\,,\nn\\
& U(\theta)\widetilde{h}(t,f)U(\theta)^{-1}=e^{-2\theta}p^2+\frac{f^2}{2\omega^2}\,\cos(2\omega t)+\frac{f^2}{2\omega^2}
\nn\end{align} for all $\theta\in S_{\theta_0}$, $f\geq0$ and $t\in\mathbb{R}$.

\begin{defi}\label{defi:resonanceAC}{\rm(\cite[Definition 1.10]{HeRa})}\\
Let $K(f,\theta)$ be given by {\rm\eqref{K}} and $f\geq0$, $\theta\in S_{\theta_0}$ with $\textrm{{\rm Im\,}}\theta>0$. The nonreal eigenvalues (which by construction are in $\mathbb{C}_-$) of $K(f,\theta)$
are defined to be the resonances of the family $\{H_\varphi(t,f)\}_{t\in\mathbb{R}}$.
\end{defi}

\begin{theorem}\label{theorem:AC}{\rm(\cite[Theorem 1.11]{HeRa})}\\
Let $\omega>0$. Let $\theta_0\in(0,\frac{\pi}{2})$ and  $\varphi\in\mathcal{D}_{\theta_0}$. Assume in addition that there exists an $f_0>0$ sufficiently small such that for all $\beta\in[-f_0,f_0]$ and all $\theta\in S_{\theta_0}$ the function $e^{f_0|x|\omega^{-1}\sin\theta_0}\varphi(e^{\theta}x+2\beta\omega^{-2})$ $(x\in\mathbb{R})$ is analytic in the parameter $\theta$ and \begin{align}\|e^{f_0|\,\cdot\,|\omega^{-1}\sin\theta_0}\varphi(e^{\theta}\cdot+2\beta\omega^{-2})\|_2\leq C\label{phi_uniformly}\end{align} for some constant $C<\infty$, uniformly for $\langle\theta,\beta\rangle\in S_{\theta_0}\times[-f_0,f_0]$. Let $H_\varphi(t,f)$ $(f\geq0,\ t\in\mathbb{R})$ be given by {\rm\eqref{H(f)_AC}} and $K(f,\theta)$ $(f\geq0,\ \theta\in S_{\theta_0})$ by {\rm\eqref{K}}.\\
Fix $\theta\in S_{\theta_0}$ with $\textrm{{\rm Im\,}}\theta>0$. Suppose there exists an eigenvalue $r_0$ (not necessarily close to 1) of $K(0,\theta)$ of multiplicity $m$. Then for all $f>0$ sufficiently small there are exactly $m$ eigenvalues (counting multiplicities) of $K(f,\theta)$ close to $r_0$ and they all converge to $r_0$ as $f\downarrow0$.
\end{theorem} The proof is by showing that $K(f,\theta)$ converges in norm resolvent sense to $K(0,\theta)$ as $f\downarrow0$ (which needs the uniformity \eqref{phi_uniformly}). For details we refer to \cite{HeRa}.

\section{Numerical example for the DC Stark case}\label{section:4}

\begin{figure}[h!]
                \centering
                \includegraphics[angle=0,width=0.5\textwidth]{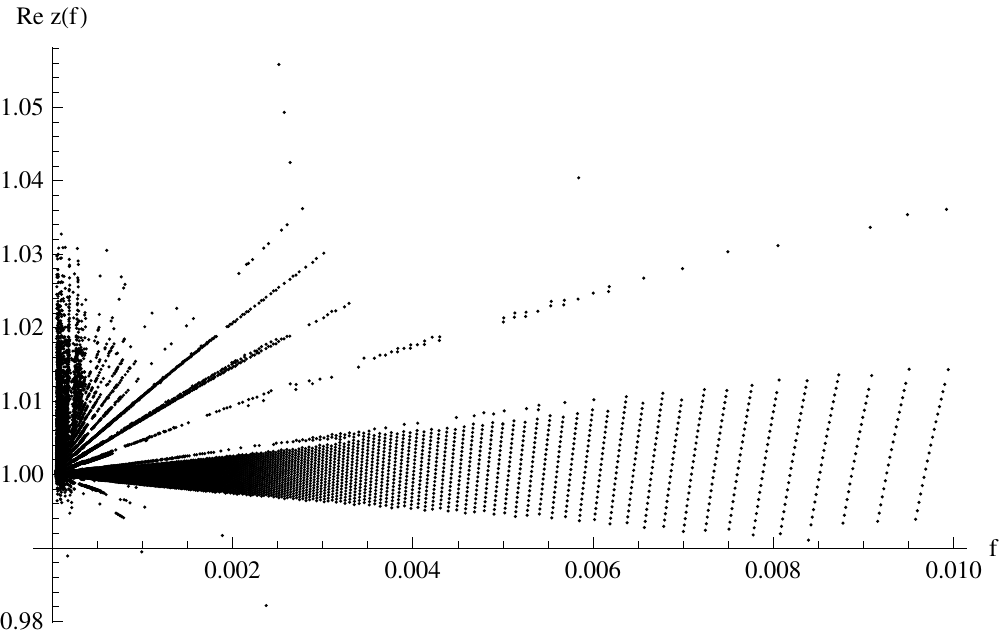}
                \caption{\normalsize Real part of resonances of $H_\varphi(f)$}\label{fig:Re_r(f)_1}\end{figure}

\begin{figure}[h!]
                \centering
                \includegraphics[angle=0,width=0.5\textwidth]{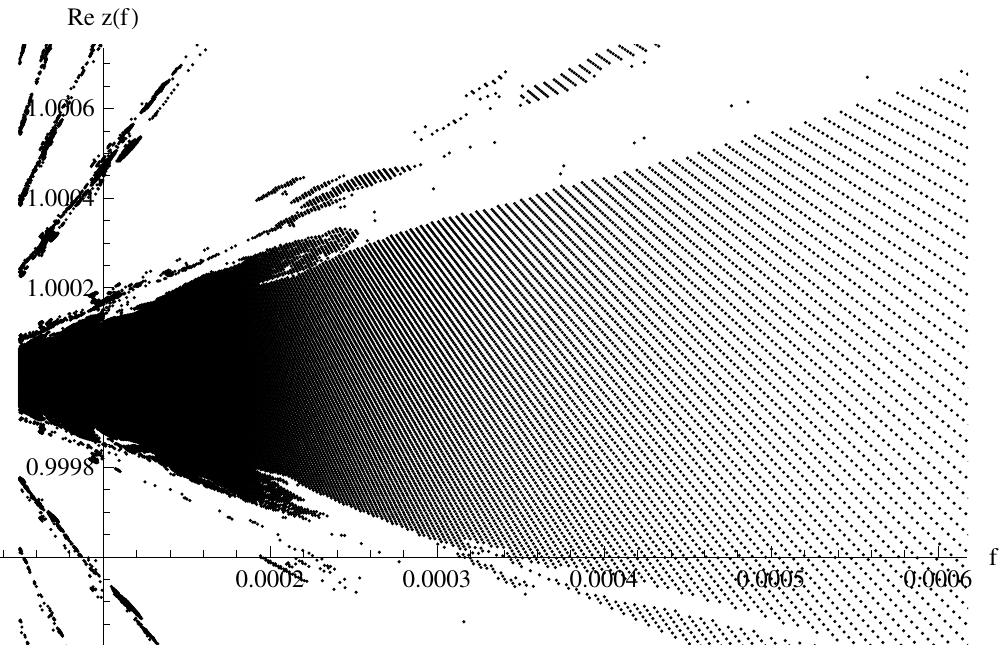}
                \caption{\normalsize Real part of resonances of $H_\varphi(f)$ on a finer scale}\label{fig:Re_r(f)}\end{figure}

\begin{figure}[h!]
                \centering
                \includegraphics[angle=0,width=0.5\textwidth]{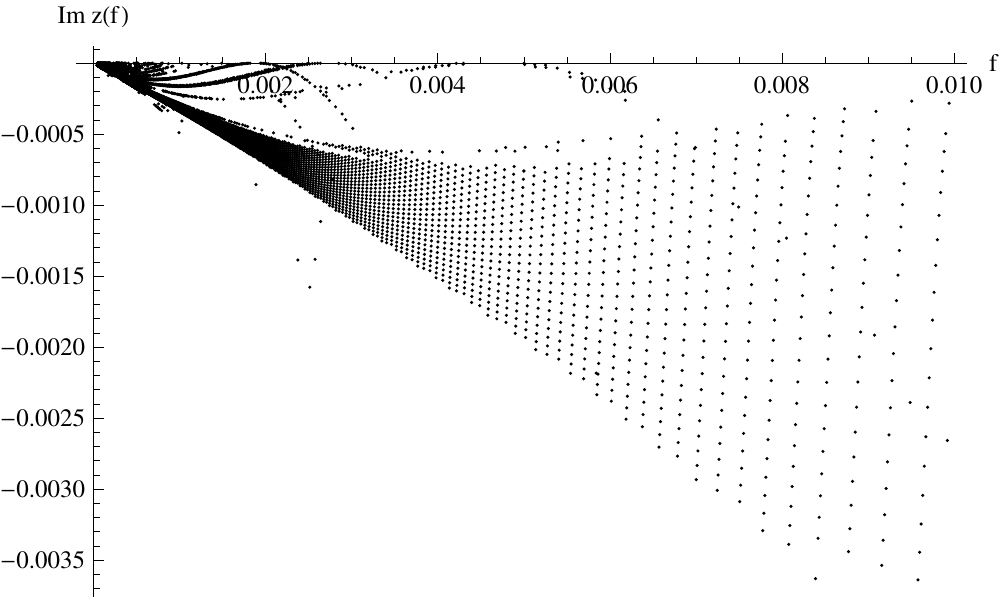}
                \caption{\normalsize Imaginary part of resonances of $H_\varphi(f)$}\label{fig:Im_r(f)_1}
\end{figure}

\begin{figure}[h!]
                \centering
                \includegraphics[angle=0,width=0.5\textwidth]{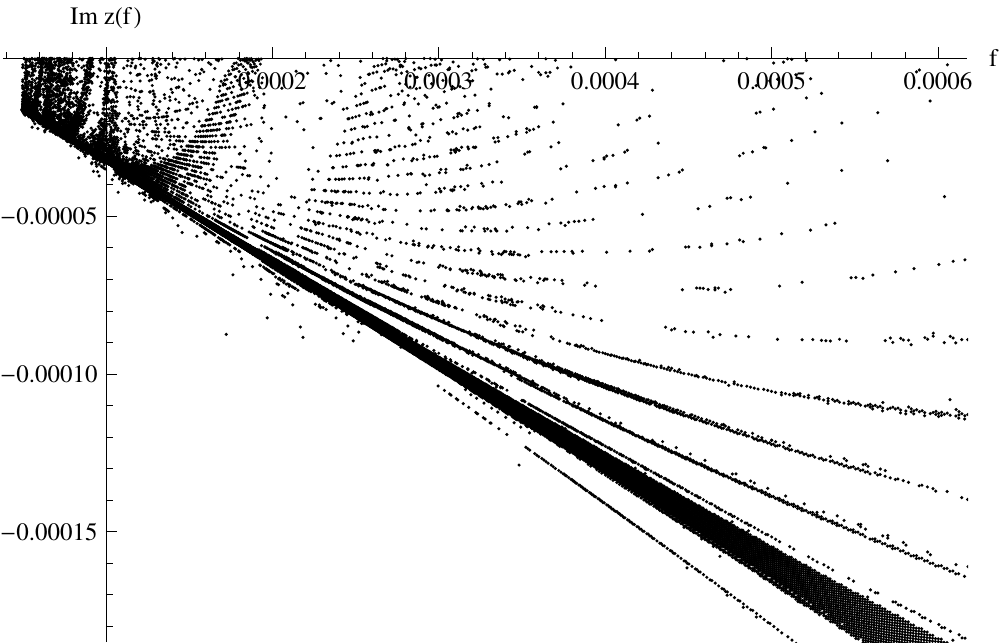}
                \caption{\normalsize Imaginary part of resonances of $H_\varphi(f)$ on a finer scale}\label{fig:Im_r(f)}
\end{figure}

In this section we present a numerical analysis of resonances of the DC Stark Hamiltonians $H_\varphi=H_\varphi(0)$ and $H_\varphi(f)$, given by \eqref{H} and \eqref{H(f)}, where
\begin{align}\varphi(x):=\frac{1}{10}\,e^{-x^2/2}\hspace{0.5cm}(x\in\mathbb{R}).\nn\end{align} The data are obtained by numerically solving \eqref{eq_zero} with Mathematica 8. The numerical results, including the figures, of this section are taken from \cite{HeRa}. For error estimates and more details we refer to \cite[Section 3]{HeRa}.
The numerically computed pre-existing resonance of $H_\varphi(0)$ is $r_0=1.01905- 0.0111115\,i$. One finds many resonances of $H_\varphi(f)$, $f>0$, near $r_0$. Figure \ref{fig:Re_r(f)_1} and Figure \ref{fig:Re_r(f)} show their real part for $f>0$ small. This real part seems to accumulate at $1$, the embedded eigenvalue of $H_0$. Note that we have no analytic proof of such a convergence result. Figure \ref{fig:Im_r(f)_1} and Figure \ref{fig:Im_r(f)} show the imaginary part of these resonances, for $f>0$ small. They all seem to converge to $0$, as they should  according to Theorem \ref{theorem:resonances_instable}.
\enlargethispage{.5cm}

${}$\\
\noindent\textbf{Acknowledgements:}\quad We thank Markus Klein for encouraging us to write this review article.

\end{document}